# Autofocus Correction of Azimuth Phase Error and Residual Range Cell Migration in Spotlight SAR Polar Format Imagery

Xinhua Mao, Daiyin Zhu, Zhaoda Zhu

(College of Electronic and Information Engineering, Nanjing University of Aeronautics and Astronautics, 210016, Nanjing, China.    Email: xinhua@nuaa.edu.cn )

**ABSTRACT**

Synthetic aperture radar (SAR) images are often blurred by phase perturbations induced by uncompensated sensor motion and /or unknown propagation effects caused by turbulent media. To get refocused images, autofocus proves to be useful post-processing technique applied to estimate and compensate the unknown phase errors. However, a severe drawback of the conventional autofocus algorithms is that they are only capable of removing one-dimensional azimuth phase errors (APE). As the resolution becomes finer, residual range cell migration (RCM), which makes the defocus inherently two-dimensional, becomes a new challenge. In this paper, correction of APE and residual RCM are presented in the framework of polar format algorithm (PFA). First, an insight into the underlying mathematical mechanism of polar reformatting is presented. Then based on this new formulation, the effect of polar reformatting on the uncompensated APE and residual RCM is investigated in detail. By using the derived analytical relationship between APE and residual RCM, an efficient two-dimensional (2-D) autofocus method is proposed. Experimental results indicate the effectiveness of the proposed method.

## I. Introduction

Synthetic aperture radar (SAR) is a coherent imaging system, which can provide high azimuth resolution by coherently processing multiple echo pulses. The coherent processing requires accurate measurements of the relative geometric relationship between radar's flight path and the scene being imaged. This geometry information can be typically provided by motion sensor such as inertial measurement unit (IMU) and global positioning system (GPS). However, these sensors can be very expensive and often difficult to provide very accurate measurements to support the ultra-high resolution imaging. To make matters worse, signal propagation through turbulent media will become another critical limiting factor as the resolution becomes finer [1]. Consequently, signal based motion compensation, i.e., autofocus algorithm, is often indispensable, and provides a necessary supplement to the IMU/GPS device, especially for very fine resolution airborne SAR processing.

The range measurement error imposes two effects on the echoes, i.e., it introduces azimuth phase error (APE) and residual range cell migration (RCM). The APE makes image defocused in the azimuth dimension, while residual RCM introduces 2-D defocus. In the phase history domain, there exists a simple linear relationship between these two error terms. That is, the APE is the multiplication of residual RCM by $4\pi/\lambda$ (where $\lambda$ is the wavelength). For radar operated with submeter wavelength ($4\pi/\lambda$ has a large value), APE will always have much more serious impact on image formation than residual RCM does. When the range measurement error is relatively small, e.g., within a range resolution cell, residual RCM effect can be neglected, and then only the APE is necessary to be compensated. This is the general presumption for almost all the existing autofocus algorithms, such as Mapdrift (MD) [2-3], Phase Difference (PD) [4], Phase Gradient Autofocus (PGA) [5-6], Eigenvector Method [7], Rank One Phase Estimate (ROPE) [8] and etc. However, as the SAR resolution becomes finer, the increased synthetic aperture length makes the accumulated range measurement error become large. On the other hand, range resolution cell becomes smaller. Therefore, residual RCM exceeding range resolution cell will become inevitable in future ultra-high resolution SAR system [9]. In this situation, to get a refocused image, not only APE, but also residual RCM, should be estimated and compensated.

To correct for the APE and residual RCM, two alternative strategies are possible. One is to estimate and compensate for the two error terms in parallel [10-13], where the APE can be extracted using conventional autofocus method, and RCM can be estimated by pulse or subaperture correlation. These two processes are performed independently. The other one works in cascaded mode [9], which is to estimate one of the error terms firstly, and then calculate the other error term from the estimated term by exploiting their analytical relationship. Typically, one can estimate the APE, and then compute the residual RCM. Compared with the first method, the latter one possesses two obvious advantages. First, the direct computation of RCM makes it more computationally efficient since no additional RCM estimation process is required. Second, the accurate estimate of APE obtained by existing autofocus method can make sure that the computed residual RCM has extreme accuracy. Due to these reasons, the cascaded correction method will be more popular in actual application.

In the cascaded correction method, a key problem is the derivation of the analytical relationship between APE and residual RCM. It is well known that there is linear relationship between APE and residual RCM in the phase history domain. However, after image formation, this simple linear relationship often doesn't exist anymore. Instead, it becomes a more complicated nonlinear mapping relationship, whose analytical expression depends on the specific image formation algorithm. In SAR literatures, many image formation algorithms have been proposed to process the raw spotlight SAR data, e.g., polar format algorithm (PFA) [14-16], chirp scaling algorithm (CSA) [17], range migration algorithm (RMA) [18-19] and backprojection algorithm (BPA) [20].

However, little work, except for works in [9, 21], has been done to analyze the uncompensated motion error after processing by image formation algorithms. Literatures [9, 21] investigated the effect of polar format resampling in PFA on uncompensated motion error, and make the cascaded correction become possible. Nevertheless, these analyses still rely upon some specific assumptions, e.g., [9]'s analysis, to simplify the derivation, uses some approximations and can only operates in broadside spotlight SAR mode, while [21] assumes that the error terms are quadratic. As a consequence, these analyses are limited to some specific applications.

In this paper, we provide a further insight into the underlying mathematical mechanism of polar reformatting in PFA. Based on this understanding, we analyze the effect of range and azimuth resampling on the uncompensated phase error, and get an accurate and more general analytical relationship between APE and residual RCM. Using this relationship, an accurate 2-D autofocus method, which can compensate for the APE and residual RCM simultaneously, is proposed.

This paper is organized as follows. In Section II, a further insight into the range and azimuth resampling of PFA is provided. Based on this new formulation approach, analysis on the residual errors in the framework of PFA is detailed in Section III. In Section IV, a 2-D autofocus method which exploits the analytical relationship between APE and residual RCM is proposed. Finally, in Section V, experimental results demonstrate the effectiveness of the proposed new autofocus approach. Section VI is the concluding remarks.

## II. Analytical Formulation of Polar Format Algorithm

For PFA, almost all the efforts are focused on the implementation of polar format transformation. However, in this paper, instead of the implementation, we will exploit the analytical formulation in detail. The simple motivation of this choice is to benefit our analysis in the next section.

*A. Signal Model*

Consider a spotlight-mode SAR operating with the geometry depicted by Fig.1. The coordinate of a generic stationary target in the illuminated scene in *XOY* is $(x_p, y_p)$. Let $t$ represent the slow time. The distance between the antenna phase center (APC) and the scene center (Point *O*) is $\mathbf{r}_c \equiv r_c(t)$, which along with the instantaneous squint angle $\mathbf{\theta} \equiv \theta(t)$ and the incidence angle $\mathbf{\varphi} \equiv \varphi(t)$ determines the instantaneous coordinate $(\mathbf{x_a}, \mathbf{y_a}, \mathbf{z_a}) \equiv [x_a(t), y_a(t), z_a(t)]$ of the APC. Note that the bold face variables in this paper are all functions of slow time $t$. Denote $\theta_{ref}$ as the reference squint angle, and $\varphi_{ref}$ the reference incidence angle. The equations in the paper can be simplified by defining $t = 0$ when $\mathbf{\theta} = \theta_{ref}$.

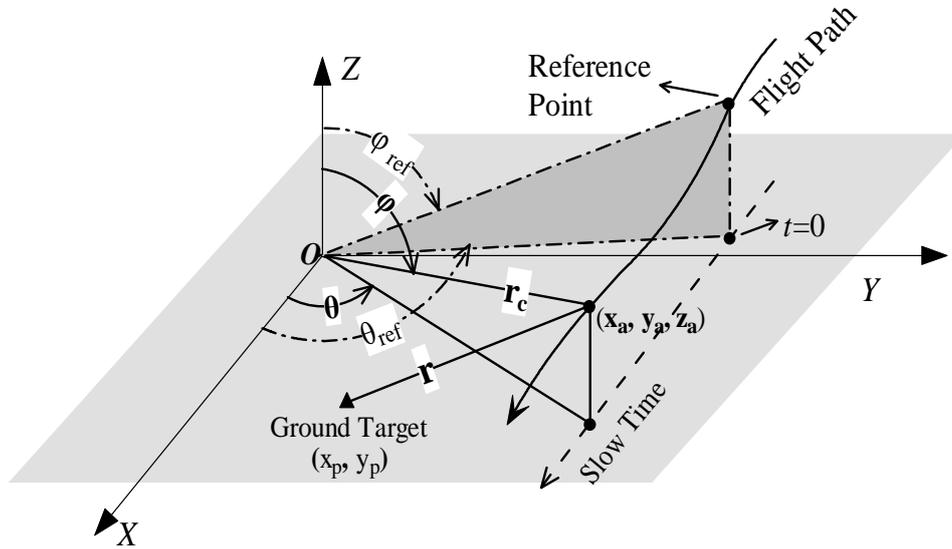

Fig.1. Spotlight SAR data collection geometry.

To proceed with the PFA, the radar echoes must be converted into the range frequency domain. The conversion method can vary with the operating modulation type. If the linear frequency modulated (LFM) signal is transmitted, the deramp technique effectively transforms the echoes into the range frequency domain at the receiver and meanwhile lowers the requirement on A/D sampling rate. Alternatively, Fourier transforming the directly sampled range signal does this job, which can be applied to any modulation type. The discussion following is based on the Fourier transform approach, but can be conveniently transplanted to the deramp one.

After matched filtering and motion compensation with respect to the scene center, the 2-D SAR signal can be expressed as

$$S(f_r,t) = A \cdot \exp\left\{j\frac{4\pi}{c}(f_0 + f_r)\mathbf{R}_\Delta\right\} \quad (1)$$

where $c$ is the speed of propagation, $f_0$ is the radar center frequency, $f_r$ is the range frequency, $A$ includes the nonessential factors of transmitted pulse envelop and azimuth antenna pattern, and $\mathbf{R}_\Delta$ is the *differential range*

$$\mathbf{R}_\Delta = \mathbf{r}_c - \mathbf{r} \equiv r_c(t) - r(t) \quad (2)$$

where $\mathbf{r} \equiv r(t)$ is the instantaneous distance between the APC and the target located on $(x_p, y_p)$.

The derivation of PFA is based on planar wavefront assumption. Under this approximation, the differential range can be simplified as

$$\mathbf{R}_\Delta = \sin\varphi\left(x_p \cos\theta + y_p \sin\theta\right) = \sin\varphi\left[\bar{x}_p \sin(\theta_{ref} - \theta) + \bar{y}_p \cos(\theta_{ref} - \theta)\right] \quad (3)$$

where

$$\begin{bmatrix} \bar{x}_p \\ \bar{y}_p \end{bmatrix} = \begin{bmatrix} \sin\theta_{ref} & -\cos\theta_{ref} \\ \cos\theta_{ref} & \sin\theta_{ref} \end{bmatrix} \begin{bmatrix} x_p \\ y_p \end{bmatrix} \quad (4)$$

are the rotated version of coordinates $(x_p, y_p)$.

By inserting (3) into (1), the 2-D signal can be rewritten as

$$S(f_r,t) = A \cdot \exp\left\{j\frac{4\pi}{c}(f_0 + f_r)\sin\varphi\left[\bar{x}_p \sin(\theta_{ref} - \theta) + \bar{y}_p \cos(\theta_{ref} - \theta)\right]\right\}. \quad (5)$$

Equation (5) shows that the phase history data are essentially a polar raster slice of the Fourier transform of the terrain reflectivity. To form the image, only 2-D discrete Fourier transform (IDFT) is required. To implement efficient Fourier transform, the fast Fourier transform (FFT) is often a popular choice. Unfortunately, the 2-D FFT requires uniformly spaced samples on a rectangular grid. Thus, to exploit the efficiency of 2-D FFT, the acquired polar samples of phase history must be resampled to a rectangular grid, and this is just what is done in PFA. In order to lower the computational burden, the 2-D resampling is usually decomposed into two tandem 1-D operations designated by the range and azimuth resampling, respectively.

From the viewpoint of range migration correction, we know that the range migration is due to the coupling between range frequency and azimuth time in phase history domain. Then the polar reformatting can also be interpreted as a decoupling procedure.

*B. Range Resampling*

The range resampling is dedicated to eliminating the coupling between range frequency and azimuth time in the coefficient of $\bar{y}_p$, making the coefficient only a linear function of range frequency. This procedure can be implemented in expression by performing an azimuth time dependent change-of-variable on range frequency in (5), i.e., setting $f_r = \vartheta_r(\bar{f}_r; t)$, where $\bar{f}_r$ is the new range frequency variable. This change-of-variable should make sure that the following decoupling transformation is achieved:

$$(f_0 + f_r)\sin\varphi\cos(\theta_{ref} - \theta) \xrightarrow{f_r = \vartheta_r(\bar{f}_r; t)} (f_0 + \bar{f}_r)\sin\varphi_{ref}. \tag{6}$$

From (6), we can easily get

$$\vartheta_r(\bar{f}_r; t) = \boldsymbol{\delta}_r \bar{f}_r + f_0(\boldsymbol{\delta}_r - 1) \tag{7}$$

where $\boldsymbol{\delta}_r = \dfrac{\sin\varphi_{ref}}{\sin\varphi\cos(\theta_{ref} - \theta)}$. Equation (7) shows that the range resampling is in essence a range frequency scaling transformation (scaling factor is $\boldsymbol{\delta}_r$) with an offset $f_0(\boldsymbol{\delta}_r - 1)$. This scaling transformation can be typically implemented by interpolation. As illustrated by Fig.2, the squares, equally spaced with interval $\Delta F$, represent the original samples in the range frequency. After resampling, if we hope that the samples in new range frequency are equally spaced with interval $\Delta F'$, then the interpolation positions, illustrated by the triangle, can be determined by the scaling mapping: $f_r = \boldsymbol{\delta}_r \bar{f}_r + f_0(\boldsymbol{\delta}_r - 1)$.

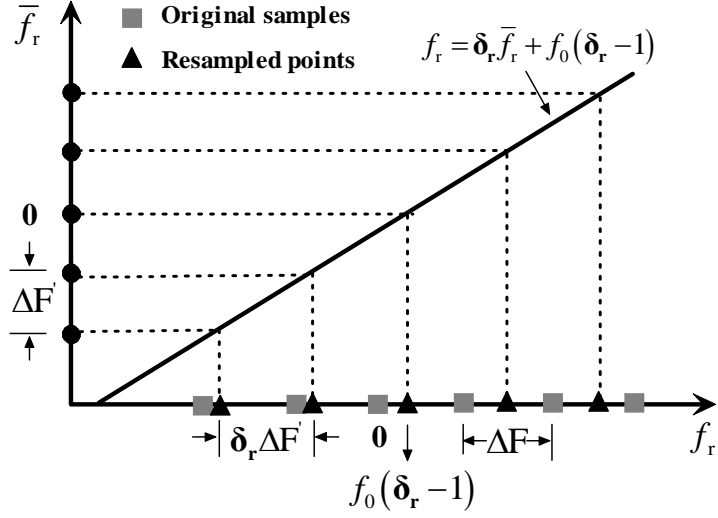

Fig.2. Illustration of range resampling in PFA.

Undergoing the above change-of-variable, the echo signal becomes

$$S_R(\bar{f}_r, t) = S[\vartheta_r(\bar{f}_r;t), t] = A \cdot \exp\left\{j\frac{4\pi}{c}(f_0 + \bar{f}_r)\sin\varphi_{ref}\left[\bar{x}_p \tan(\theta_{ref} - \boldsymbol{\theta}) + \bar{y}_p\right]\right\}. \tag{8}$$

In (8), there has been no coupling between the new range frequency $\bar{f}_r$ and the azimuth time variable $t$ in the coefficient of $\bar{y}_p$ term.

*C. Azimuth Resampling*

Similarly to range resampling, the azimuth resampling is dedicated to eliminating the coupling between azimuth time and range frequency in the coefficient of $\bar{x}_p$ in (8), making the coefficient only a linear function of azimuth time. This procedure can be implemented in expression by performing a range frequency dependent change-of-variable on azimuth time, which can be denoted as $t = \vartheta_a(\bar{t};\bar{f}_r)$. This change-of-variable should make sure that the following transformation is achieved:

$$(f_0 + \bar{f}_r)\tan(\theta_{ref} - \boldsymbol{\theta}) \xrightarrow{t=\vartheta_a(\bar{t};\bar{f}_r)} f_0\Omega\bar{t} \tag{9}$$

where $\Omega$ is a constant determined by this azimuth resampling process. If the APC travels ideally in parallel with *OX* (thus $\mathbf{y}_a$ equals a constant $y_a$) at constant forward velocity $u_X$, $\Omega$ can be typically set as

$$\Omega = \frac{u_X \sin^2\theta_{ref}}{y_a}.$$

Without loss of generality, we assume that the radar platform flies at an arbitrary path. In this situation, only the numerical relationship between $\tan(\theta_{\text{ref}} - \boldsymbol{\theta})$ and $t$, rather than the analytical one, can be provided at hand. Thereby it will be a difficult task to derive an analytical solution of $\vartheta_a(\bar{t}; \bar{f}_r)$ from (9). To benefit our analysis in the next section, we divide the azimuth resampling formulated as $\vartheta_a(\bar{t}; \bar{f}_r)$ into two cascaded resampling procedures, which are expressed as $\vartheta_{a1}(\hat{t})$ and $\vartheta_{a2}(\bar{t}; \bar{f}_r)$, respectively. Then (9) can be equivalently implemented as following

$$(f_0 + \bar{f}_r)\tan(\theta_{\text{ref}} - \boldsymbol{\theta}) \xrightarrow{t = \vartheta_{a1}(\hat{t})} (f_0 + \bar{f}_r)\Omega\hat{t} \xrightarrow{\hat{t} = \vartheta_{a2}(\bar{t}; \bar{f}_r)} f_0 \Omega \bar{t} \ . \tag{10}$$

Firstly, a range-frequency-independent azimuth time transformation, i.e., $t = \vartheta_{a1}(\hat{t})$, is performed to implement the linearization of $\tan(\theta_{\text{ref}} - \boldsymbol{\theta})$. Such a procedure in this paper is referred to as RCM linearization. Its implementation can be illustrated in Fig.3. For uniformly spaced time instants of $t$ represented by the squares on the horizontal axis, the corresponding values of $\tan(\theta_{\text{ref}} - \boldsymbol{\theta})$ are in general non-uniformly spaced for realistic motion of the APC. However, the range resampled signal in (8) can be interpolated on specific time instants represented by the triangles, on which $\tan(\theta_{\text{ref}} - \boldsymbol{\theta})$ progresses arithmetically.

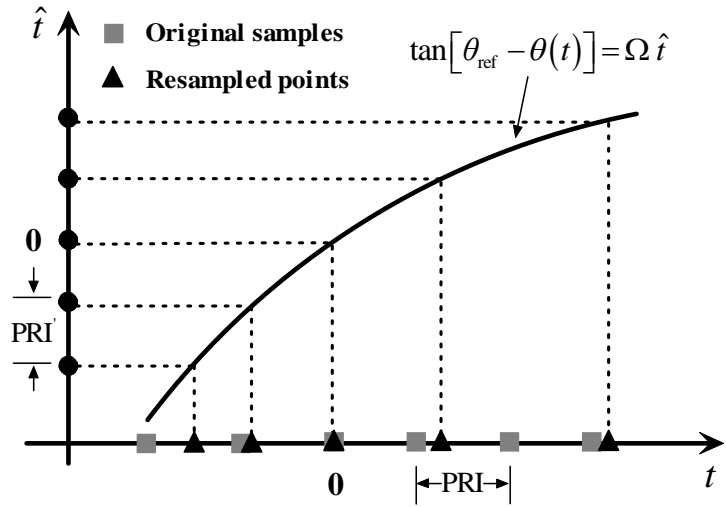

Fig.3.   Illustration of RCM linearization in PFA.

With the linearization process the $\tan(\theta_{\text{ref}} - \boldsymbol{\theta})$ term in (8) can be replaced by a linear function of $\hat{t}$.

Therefore, after RCM linearization, $S_R(\bar{f}_r, t)$ in (8) becomes

$$S_L(\bar{f}_r, \hat{t}) = S_R[\bar{f}_r, \vartheta_{a1}(\hat{t})] = A \cdot \exp\left\{j\frac{4\pi}{c}(f_0 + \bar{f}_r)\sin\varphi_{ref}(\bar{x}_p\Omega\hat{t} + \bar{y}_p)\right\}. \tag{11}$$

From (11), we can see that the coefficient of $\bar{x}_p$ is still range frequency dependent. Therefore the completely decoupling is not yet achieved.

The second resampling procedure, formulated as $\vartheta_{a2}(\bar{t}\ ;\bar{f}_r)$, is a range frequency dependent azimuth time resampling. Consulting (10), we can easily get

$$\vartheta_{a2}(\bar{t}\ ;\bar{f}_r) = \frac{f_0}{f_0 + \bar{f}_r}\bar{t}. \tag{12}$$

This is the well-known keystone transform (KT), which is used for bulk correction of arbitrary linear RCM through decoupling azimuth time and range frequency [22-23].

After KT, the signal in (11) becomes

$$S_{KT}(\bar{f}_r, \bar{t}) = S_L[\bar{f}_r, \vartheta_{a2}(\bar{t};\bar{f}_r)] = \tilde{A} \cdot \exp\left\{j\frac{4\pi}{c}\sin\varphi_{ref}(\bar{x}_p f_0\Omega\bar{t} + \bar{y}_p\bar{f}_r)\right\} \tag{13}$$

where $\tilde{A} = A \cdot \exp\left(j\frac{4\pi\sin\varphi_{ref}}{\lambda}\bar{y}_p\right)$ is a constant.

D. *2-D Fourier Transform*

In (13) there has been no coupling between azimuth time $\bar{t}$ and range frequency $\bar{f}_r$. The residual 2-D sinusoid can be converted into a focused target response via the 2-D Fourier transform with respect to $\bar{t}$ and $\bar{f}_r$, which is expressed as the product of two sinc functions:

$$\mathbf{F}[S_{KT}(\bar{f}_r, \bar{t})] = \tilde{A} \cdot \mathrm{sinc}_r\left(\tau - \frac{2\sin\varphi_{ref}}{c}\bar{y}_p\right) \cdot \mathrm{sinc}_a\left(f_t - \frac{2\Omega\sin\varphi_{ref}}{\lambda}\bar{x}_p\right) \tag{14}$$

where $\mathbf{F}[\cdot]$ represents the 2-D Fourier transform, $f_t$ is the azimuth Doppler frequency, $\tau$ is the range time, $\mathrm{sinc}_a(\cdot)$ and $\mathrm{sinc}_r(\cdot)$ are azimuth and range sinc functions, respectively, which are defined by

$$\begin{aligned}\mathrm{sinc}_a(u) &= \sin(\pi B_d u)/(\pi B_d u) \\ \mathrm{sinc}_r(u) &= \sin(\pi B_\tau u)/(\pi B_\tau u)\end{aligned} \tag{15}$$

where $B_d$ and $B_\tau$ are the azimuth signal bandwidth and range signal bandwidth, respectively.

Equation (14) is the target's impulse response (IPR) function of the PFA with far-field approximation. Now it is rather clear that the azimuth resampling procedure in PFA is essentially the combination of RCM linearization and KT. The analytical expression of $\vartheta_{a1}(\hat{t})$ may not be obtained under arbitrary radar flight path, but this will not affect our analysis. We will show, in the next section, that only the KT contributes to the derivation of analytical relationship between APE and residual RCM. Although we divide the azimuth resampling into two individual procedures to facilitate our analysis, it should be noted that in the practical implementation of PFA, still only one interpolation is required to accomplish both procedures of RCM linearization and KT.

### III. Effect of Polar Format Resampling on Uncompensated Phase Error

In the above development of PFA, we assumed that the relative range between radar and target scene is accurately known. However in practical situation, due to inaccurate motion measurement or deleterious atmospheric effects, range errors are often inevitable. Therefore, the actual differential range should be expressed as

$$\mathbf{R}_\Delta = \left[ \bar{x}_p \sin\varphi \sin(\theta_{ref} - \boldsymbol{\theta}) + \bar{y}_p \sin\varphi \cos(\theta_{ref} - \boldsymbol{\theta}) \right] + \mathbf{R}_E \tag{16}$$

where $\mathbf{R}_E \equiv R_E(t)$ represents all the range errors.

Inserting (16) into (1), we can get the actual echo signal

$$S(f_r, t) = A \cdot \exp\left\{ j\frac{4\pi}{c}(f_0 + f_r)\left[ \bar{x}_p \sin\varphi \sin(\theta_{ref} - \boldsymbol{\theta}) + \bar{y}_p \sin\varphi \cos(\theta_{ref} - \boldsymbol{\theta}) + \mathbf{R}_E \right] \right\}. \tag{17}$$

Comparing this actual signal with the expected signal of (5) in PFA, we can get the error phase term

$$\phi(f_r, t) = \frac{4\pi}{c}(f_0 + f_r)\mathbf{R}_E = \phi_0(t) + \phi_1(t) f_r \tag{18}$$

where $\phi_0(t) = \frac{4\pi}{\lambda}\mathbf{R}_E$ is the APE, and $\phi_1(t) = \frac{4\pi}{c}\mathbf{R}_E$ is related to the residual RCM. Clearly, at this moment, the APE and residual RCM have a simple linear relationship.

From the previous section, we have known that the range resampling in PFA is essentially a scaling transform in range frequency defined in (7). Therefore, after this range resampling, the signal in (17) can be expressed as

$$S_R(\bar{f}_r, t) = S[\vartheta_r(\bar{f}_r; t), t] = A \cdot \exp\left\{j \frac{4\pi(f_0 + \bar{f}_r)\sin\varphi_{ref}}{c}\left[\bar{x}_p \tan(\theta_{ref} - \mathbf{\theta}) + \bar{y}_p + \boldsymbol{\varepsilon}\right]\right\} \quad (19)$$

where $\boldsymbol{\varepsilon} = \dfrac{\boldsymbol{\delta}_r \mathbf{R}_E}{\sin\varphi_{ref}} \equiv \varepsilon(t)$.

Consulting (19) and (17), we can find that the range resampling also performs effects on the phase error. The resulted phase errors are

$$\phi(\bar{f}_r, t) = \frac{4\pi(f_0 + \bar{f}_r)}{c} \boldsymbol{\delta}_r \mathbf{R}_E = \phi_0(t) + \phi_1(t)\bar{f}_r \quad (20)$$

where $\phi_0(t) = \dfrac{4\pi}{\lambda}\boldsymbol{\delta}_r \mathbf{R}_E$ is the APE, and $\phi_1(t) = \dfrac{4\pi}{c}\boldsymbol{\delta}_r \mathbf{R}_E$ is related to the residual RCM. Apparently, the linear relationship between APE and residual RCM still holds.

The second step of PFA is azimuth resampling. To faciliate the following analysis, we have divided it into two cascaded resampling procedure, i.e., RCM linearization and KT, in the previous section. RCM linearization is a range-frequency-independent azimuth time resample formulated as $t = \vartheta_{a1}(\hat{t})$. After this RCM linearization, the signal in (19) becomes

$$S_L(\bar{f}_r, \hat{t}) = S_R[\bar{f}_r, \vartheta_{a1}(\hat{t})] = A \cdot \exp\left\{j\frac{4\pi}{c}(f_0 + \bar{f}_r)\sin\varphi_{ref}(\bar{x}_p\Omega\hat{t} + \bar{y}_p + \mathbf{\eta})\right\} \quad (21)$$

where $\mathbf{\eta} = \varepsilon[\vartheta_{a1}(\hat{t})] \equiv \eta(\hat{t})$ is the residual error corresponding to a 2-D phase error term

$$\phi(\bar{f}_r, \hat{t}) = \frac{4\pi}{c}(f_0 + \bar{f}_r)\sin\varphi_{ref}\mathbf{\eta} = \phi_0(\hat{t}) + \phi_1(\hat{t})\bar{f}_r \quad (22)$$

where $\phi_0(\hat{t}) = \dfrac{4\pi}{\lambda}\sin\varphi_{ref}\mathbf{\eta}$ and $\phi_1(\hat{t}) = \dfrac{4\pi}{c}\sin\varphi_{ref}\mathbf{\eta}$ are the APE and residual RCM, respectively. Compared with the corresponding terms before RCM linearization, it is clearly that both of APE and residual

RCM undergo a change, but in a synchronous style. Thereby their linear relationship keeps on.

The final step is to perform KT on (21), which results in

$$S_{KT}(\bar{f}_r, \bar{t}) = S_L[\bar{f}_r, \vartheta_{a2}(\bar{t};\bar{f}_r)] = \tilde{A} \cdot \exp\left\{j\frac{4\pi}{c}\sin\varphi_{ref}(\bar{x}_p f_0 \Omega \bar{t} + \bar{y}_p \bar{f}_r) + \phi(\bar{f}_r, \bar{t})\right\} \quad (23)$$

where $\tilde{A} = A \cdot \exp\left(j\frac{4\pi \sin\varphi_{ref}}{\lambda}\bar{y}_p\right)$ is a constant, and $\phi(\bar{f}_r, \bar{t}) = \frac{4\pi(f_0 + \bar{f}_r)\sin\varphi_{ref}}{c}\eta\left(\frac{f_0}{f_0 + \bar{f}_r}\bar{t}\right)$ is the

2-D phase error after KT. To facilitate the analysis of error effect, Taylor series expansion of phase error with respect to range frequency is performed

$$\phi(\bar{f}_r, \bar{t}) = \phi_0(\bar{t}) + \phi_1(\bar{t})\bar{f}_r + \phi_2(\bar{t})\bar{f}_r^2 + \cdots \quad (24)$$

where the first term in the right hand is the APE term, the second term is related to the residual RCM [the residual RCM is $\frac{c}{4\pi}\phi_1(\bar{t})$], and these two terms are the main limiting factors to target focusing. Higher order terms in general are not significant in most radar specific applications and often ignored. By simple derivation, the first two Taylor coefficients in (24) can be derived as following

$$\phi_0(\bar{t}) = \phi(\bar{f}_r, \bar{t})\big|_{\bar{f}_r=0} = \frac{4\pi \sin\varphi_{ref}}{\lambda}\eta(\bar{t}) \quad (25)$$

$$\phi_1(\bar{t}) = \frac{d\phi(\bar{f}_r, \bar{t})}{d\bar{f}_r}\bigg|_{\bar{f}_r=0} = \frac{4\pi \sin\varphi_{ref}}{c}\left[\eta(\bar{t}) - \bar{t}\cdot\frac{d\eta(\bar{t})}{d\bar{t}}\right]. \quad (26)$$

Consulting (25) and (26), it is easy to get the analytical relationship between $\phi_0(\bar{t})$ and $\phi_1(\bar{t})$:

$$\phi_1(\bar{t}) = \frac{1}{f_0}\left\{\phi_0(\bar{t}) - \bar{t}\cdot\frac{d\phi_0(\bar{t})}{d\bar{t}}\right\}. \quad (27)$$

If we let $r_E(\bar{t})$ represent the residual RCM and $\phi_E(\bar{t})$ the APE, then their analytical relationship is

$$r_E(\bar{t}) = \frac{\lambda}{4\pi}\left\{\phi_E(\bar{t}) - \bar{t}\cdot\frac{d\phi_E(\bar{t})}{d\bar{t}}\right\}. \quad (28)$$

From the above analysis, we can know that both range resampling and RCM linearization perform synchronous changes on the structure of APE and residual RCM, thus they can keep the linear relationship between APE and residual RCM all the time. However, unlike the above two procedures, KT imposes different effects on the APE and residual RCM. It makes the intrinsic linear relationship don't exist any more. Instead is a nonlinear relationship, which has been shown in (28). Fortunately, this nonlinear relationship is flight path independent. This makes possible an efficient cascaded correction of APE and residual RCM.

## IV. Autofocus Correction of APE and Residual RCM in PFA Imagery

From analysis in the previous section, we know that there exists a flight-path-independent analytical relationship between APE and residual RCM in the defocused PFA imagery. Consequently, once one of the two error terms has been estimated, then the other term can be computed directly by exploiting their analytical relationship. Typically, we can estimate the APE firstly, and then compute the residual RCM from this estimated APE. The motivation of this choice is twofold. First, almost all the efforts in SAR autofocus literatures are focused on the APE estimation, and there have been many established APE estimation algorithms which can provide phase estimation with high accuracy. Secondly, compared with the APE, the residual RCM is much less sensitive to error source. Therefore, if the accuracy of APE estimation can satisfy the requirement of phase compensation, then the computed residual RCM will definitely possess a high accuracy with respect to RCM correction.

Based on the above consideration, an autofocus method used for cascaded estimation and compensation of APE and residual RCM is proposed, whose processing flow is illustrated in Fig.4. This cascaded correction method includes two key parts, i.e., APE estimation and residual RCM computation.

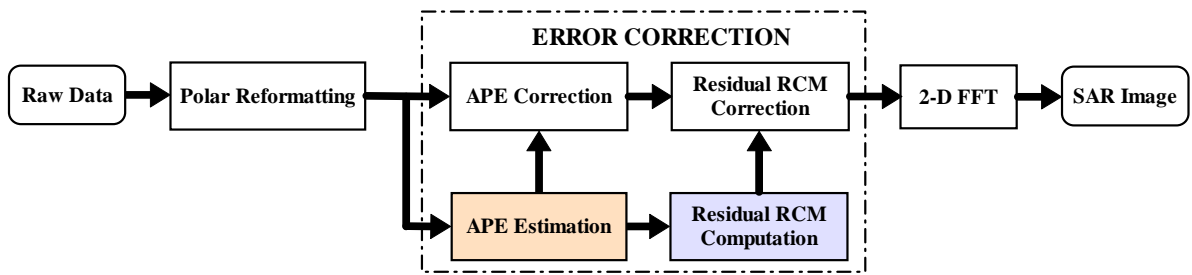

Fig.4. Flow diagram of 2-D autofocus algorithm in the framework of PFA.

*A. APE Estimation*

APE estimation can typically be implemented using conventional autofocus algorithm. But a necessary modification will be required when residual RCM exceeds range resolution cell, since this is not taken into account in conventional autofocus method. To solve this problem, at least two alternative strategies at hand can provide this capability.

The most straightforward way is to perform a preprocessing on the data to reduce the range resolution, thereby keeping the residual RCM not to exceed a coarse range resolution cell. After this preprocessing, APE can be estimated by conventional autofocus techniques such as PGA. However, this resolution-reduction preprocessing will also incur an inherent drawback, that is, the number of independent range samples used to estimate APE is decreased. The larger the residual RCM is, the fewer the independent range samples are. When the reduced independent range sample number is not large enough, the APE estimation will suffer from performance degradation.

In such situation, subaperture autofocus method may be more appropriate. In subaperture method, the long coherent processing interval (CPI) can be first divided into several subapertures. As long as the length of each subaperture is small enough, the residual RCM in subaperture will always be negligible. Therefore, in each subaperture, traditional autofocus method, e.g., PGA in this paper, can be used to extract the subaperture phase error (SPE). However, subaperture autofocusing also incurs the problem of coherent SPE combination, arising from the fact that traditional autofocus method fails to reconstruct the linear component of SPE. Although the linear phase error has no impact on the focal quality of subaperture image, its variation in inter-subapertures will prohibit the coherent combination of SPE. For example, we assume a phase error in the full aperture is illustrated in Fig.5(a). We divide it into five subapertures. Generally, there are different linear phase terms in different subapertures, and this can be seen in Fig.5(b). If the linear phase error term in each subaperture is eliminated, then a direct combination of residual subaperture phase errors will result in Fig.5(c), from which we can see a distinct discontinuity in the boundary of subaperture. To combine the SPE coherently, an additional estimation of linear phase error, or at least the relative linear phase error (RLPE) between subapertures will be required.

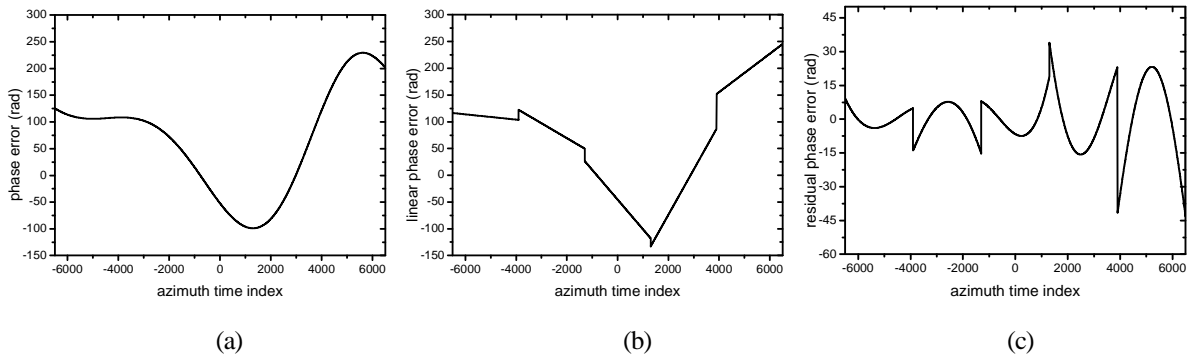

(a) (b) (c)

Fig.5. Combination of subaperture phase error (a) full-aperture phase error; (b) linear phase term in each subaperture; (c) high order phase terms in each subaperture.

It is well-known that linear phase error induces an azimuth shift in the image domain. Therefore, if we can detect the relative azimuth shift between subaperture images, then the RLPE can be estimated. Typically, relative azimuth shift can be estimated via cross-correlating the subaperture image pair in azimuth dimension. Using the RLPE estimate, the SPE can then be combined coherently. Fig.6 shows this APE estimation method in detail (Its performance analysis can be seen in [24]).

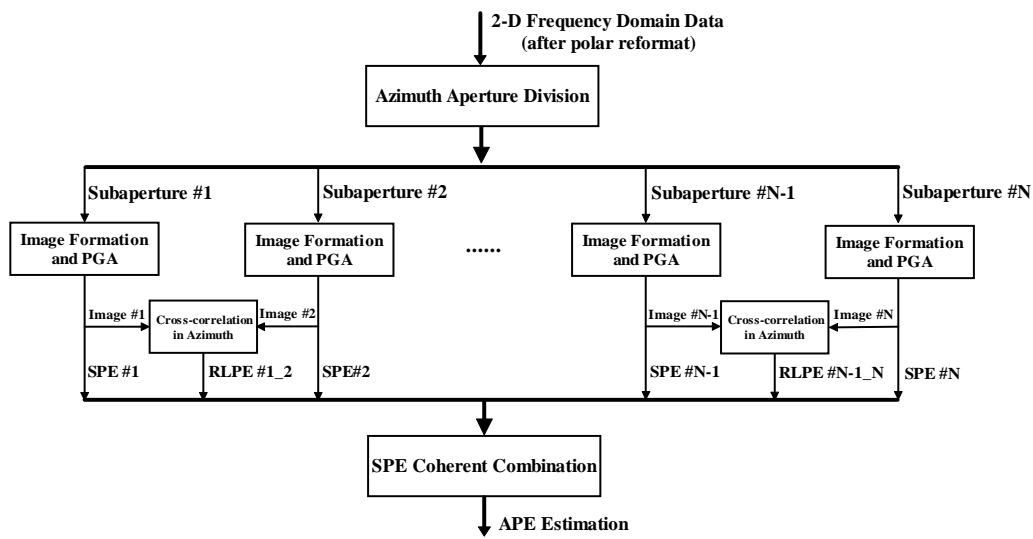

Fig.6. Flowchart of the subaperture autofocus method.

*B. Residual RCM Computation*

The second key part is the computation of residual RCM, which is based on exploiting the analytical relationship between APE and residual RCM. Although it is well known that the azimuth phase and RCM have a linear relationship in the echo data (also referred to as phase history domain), after image formation processing (polar format transformation in this work), this simple relationship between APE and residual RCM doesn't exist anymore. Instead, a more complicated nonlinear relationship, e.g., eq. (28) in PFA, is present. Fortunately, from (28) we can see that the mapping from APE to residual RCM is one-to-one and flight-path independent. Thereby, once the APE is estimated, the corresponding residual RCM can be calculated from the estimated APE directly. This direct computation of residual RCM eliminates an additional independent estimation process, thereby possessing a high computational efficiency.

*C. Effect of APE Estimation Error on Computed Residual RCM*

Since the residual RCM is computed from the estimated APE, then the estimation error of residual RCM will depend on not only the analytical relationship between residual RCM and APE, but also the estimation error of APE. For APE estimation, there are mainly two error sources which should be taken into account. One is the linear phase error, which is inherent in traditional autofocusing methods since they are aimed at refocusing the defocused image caused by high order phase error. Although linear term in APE has no impact on the refocus of image, it has the possibility to introduce additional migration into the computed residual RCM. The second is the high frequency random error. This error is relatively small, and often incurs a negligible sidelobe effect. However, if not eliminated, it will perform a more serious impact on the computed residual RCM, and this will be clearly seen in the following discussion.

To analyze the effect of linear error in APE on the computed residual RCM, without loss of generality, we assume that the linear term $\hat{\phi}_{\text{linear}}(\bar{t}) = a\bar{t}$, where $a$ is an arbitrary constant. Consulting (28), the corresponding

computed residual RCM is

$$\hat{r}_{\text{linear}}(\bar{t}) = \frac{\lambda}{4\pi}\left\{\hat{\phi}_{\text{linear}}(\bar{t}) - \bar{t} \cdot \frac{\mathrm{d}\hat{\phi}_{\text{linear}}(\bar{t})}{\mathrm{d}\bar{t}}\right\} = 0. \quad (29)$$

This is an exciting conclusion, which shows that no additional migration will be introduced into the computed residual RCM even if APE has a linear phase offset. This important benefit is due to the underlying keystone transform in polar reformatting.

To analyze the effect of random phase error in APE on the computed residual RCM, it is convenient to express the analytical relationship between APE and residual RCM in discrete form, i.e.,

$$\hat{r}_{\text{E}}(m) = \frac{\lambda}{4\pi}\left\{\hat{\phi}_{\text{E}}(m) - m \cdot \left[\hat{\phi}_{\text{E}}(m) - \hat{\phi}_{\text{E}}(m-1)\right]\right\}, \quad -\frac{M}{2} \leq m < \frac{M}{2} \quad (30)$$

where $m$ is azimuth time index and $M$ is the aperture length.

We assume that the APE estimation has a white-noise error, whose magnitude is identified by the variance of APE, i.e., $\sigma_\phi^2 = \text{var}\left[\hat{\phi}_{\text{E}}(m)\right]$. Then from (30), we can derive that the variance of the computed residual RCM is

$$\sigma_r^2(m) = \text{var}\left[\hat{r}_{\text{E}}(m)\right] = \left(\frac{\lambda}{4\pi}\right)^2 \left(1 - 2m + 2m^2\right)\sigma_\phi^2, \quad -\frac{M}{2} \leq m < \frac{M}{2}. \quad (31)$$

Unlike the variance of APE, we can see from (31) that the variance of computed residual RCM is dependent on the azimuth time. At the edge of aperture (where $|m|$ is large), the variance becomes large. To show this effect intuitively, a numerical simulation is performed. We assume that a relatively small white-noise phase error is added into the APE. Fig. 7(a) shows the APE with and without phase noise, respectively. By exploiting (30), we can compute the corresponding residual RCM, which is shown in Fig.7 (b). From this figure, it can be clearly seen that the random error is enlarged in the computed residual RCM, especially at the place far from aperture center. Apparently, this contaminated RCM has to be improved when used to correct for residual RCM in PFA imagery.

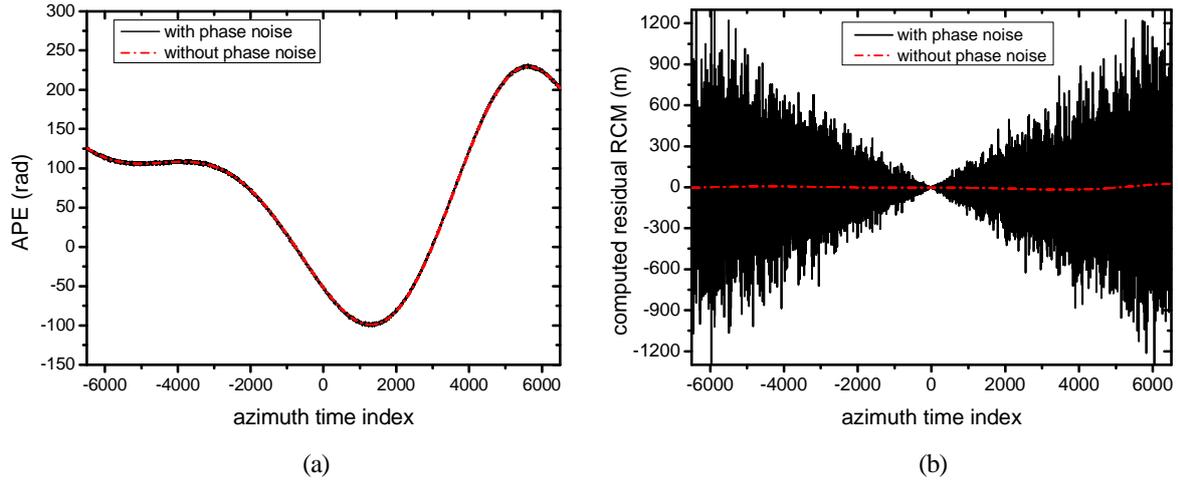

Fig.7. Effect of phase noise on (a) the APE and (b) the corresponding residual RCM.

Fig.8 (a) and (b) show the frequency spectrum of the APE and residual RCM, respectively. Fortunately, in the frequency domain, the random phase noise is mainly related to the high-frequency components. While in most cases, the residual RCM we are concerned about often has low-frequency property. Therefore, to reduce the effect of random phase noise on the computed residual RCM, low-pass filtering can be performed on the APE and/or the computed residual RCM. In this example, we perform low-pass filtering on the APE. After this filtering, the computed residual RCM and its frequency spectrum are illustrated in Fig.9 (a) and (b), respectively. It is clear that the effect of phase noise has been reduced to a tolerable level, i.e., the residual range perturbation is less than a range resolution cell.

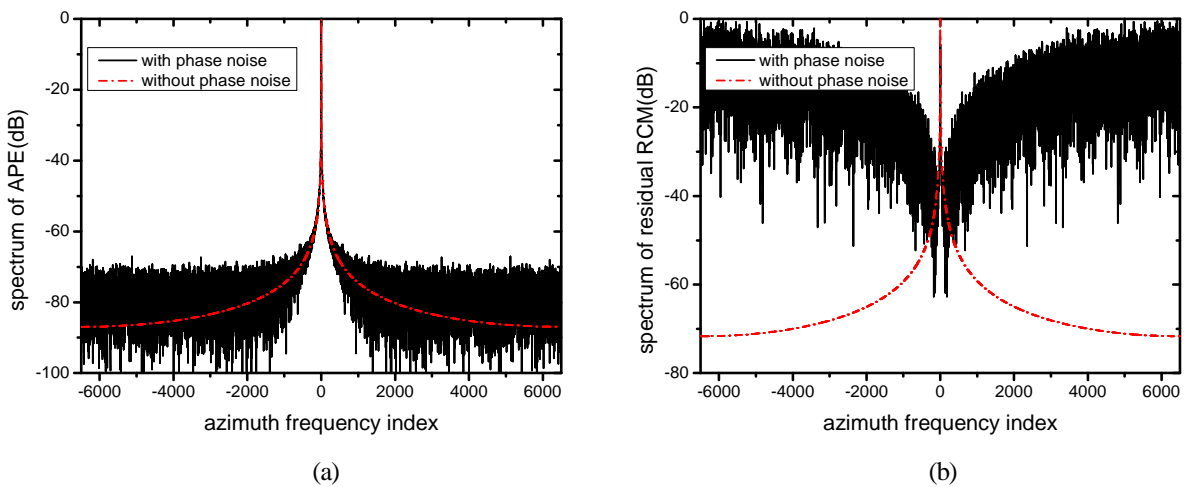

Fig.8. Spectrum of APE and computed residual RCM.

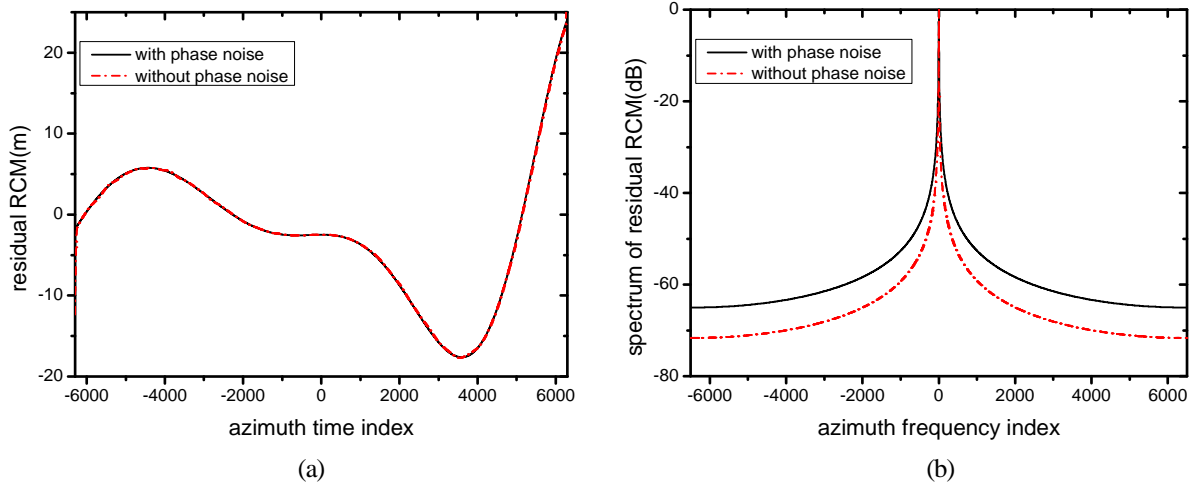

Fig.9. Results after low-pass filtering of APE. (a) residual RCM; (b) spectrum of residual RCM.

*D. Iteration*

From the above analysis, we know that the estimation of APE and that of residual RCM are mutually interacting. That is, accurate measurement of APE is limited by the residual RCM because the error energy is spread across several range resolution cells, while estimation error of APE, on the contrary, also performs impact on residual RCM estimation since the residual RCM in our method is calculated from the APE estimation. Although some schemes, e.g., reduction of range resolution, low-pass filtering of the APE, and etc., can be used to attenuate these interactions, there still exist the cases where residual interaction can't be ignored. In this situation, it may be necessary to execute the estimation and correction of APE and residual RCM in an iterative manner. That is, after the image is corrected using the initial estimation of APE and residual RCM, the entire process is repeated on this refocused image. Our experience has shown that 2-3 iterations will provide satisfactory results.

## V. Experimental Results

*A. Simulation Example*

Simulation experiment is performed to verify the theoretical analysis in Section III. Table I lists the main parameters used in the simulation. The low center frequency, short slant range, and large integration angle are very unlikely to occur simultaneously in reality, but are herein assumed in order to clearly identify the variation of APE and RCM trajectory in the different processing stages of the PFA.

Table I PARAMETERS FOR THE SIMULATION

| Parameter Description | Value |
|---|---|
| Radar wavelength | 0.6 m |
| Bandwidth of transmission signal | 500 MHz (range resolution 0.3 m) |
| Azimuth integration angle | 57.3 deg (azimuth resolution 0.3 m) |
| Scene center range at aperture center | 5000 m |
| Platform altitude | 3000 m |
| Nominal radar forward velocity | 100 m/s |
| Reference squint angle | 75 deg |
| Point target location | (0 m, 0 m) |

A nominal linear flight path is assumed in the polar format image formation. But in reality, deviations from this nominal linear trajectory in Y and Z direction are introduced, which are shown in Fig.10. A ground point target is assumed to be located in the scene center. Therefore after motion compensation to the scene center, all the phase terms in the echo signal are related to the errors induced by the radar's maneuver.

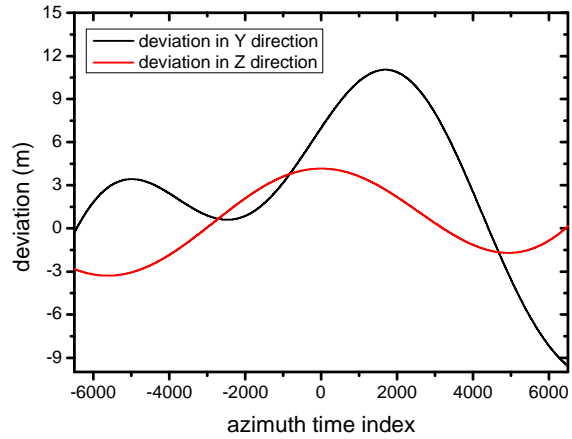

Fig.10.  Deviations of radar platform from the nominal flight path.

Polar format transform is essentially a resampling procedure. We know that resampling satisfies the property: $\mathbf{P}\left[e^{\phi(f_r,t)}\right] = e^{\mathbf{P}\left[\phi(f_r,t)\right]}$, where $\mathbf{P}[\cdot]$ represents the resample operator and $\phi(f_r,t)$ is the 2-D phase. Thus, to get the effect of polar format transform on the 2-D echo phase, we can also perform polar format transform directly on the phase terms of echo signal, instead of the echo signal itself. This change will facilitate our analysis on the APE and residual RCM. For example, the APE can be directly obtained by evaluating 2-D error phase term $\phi(f_r, t)$ at $f_r = 0$, and residual RCM by $\frac{c}{4\pi} \frac{\mathrm{d}\phi(f_r, t)}{\mathrm{d}f_r}\bigg|_{f_r=0}$, since no phase unwrapping is required. Using this method, the measured APE and residual RCM in the different processing stages of the PFA are depicted in Fig.11(a) and Fig.11(b), respectively. Comparative analysis on these two figures shows that a linear relationship between APE and residual RCM exists in all stages before KT. However, the KT breaks down this simple relationship. To verify the correctness of their analytical relationship shown in (28), we also give, in Fig.11(b), the residual RCM computed from APE by using (28). It can be clearly seen that the computed residual RCM matches quite well with the measured residual RCM.

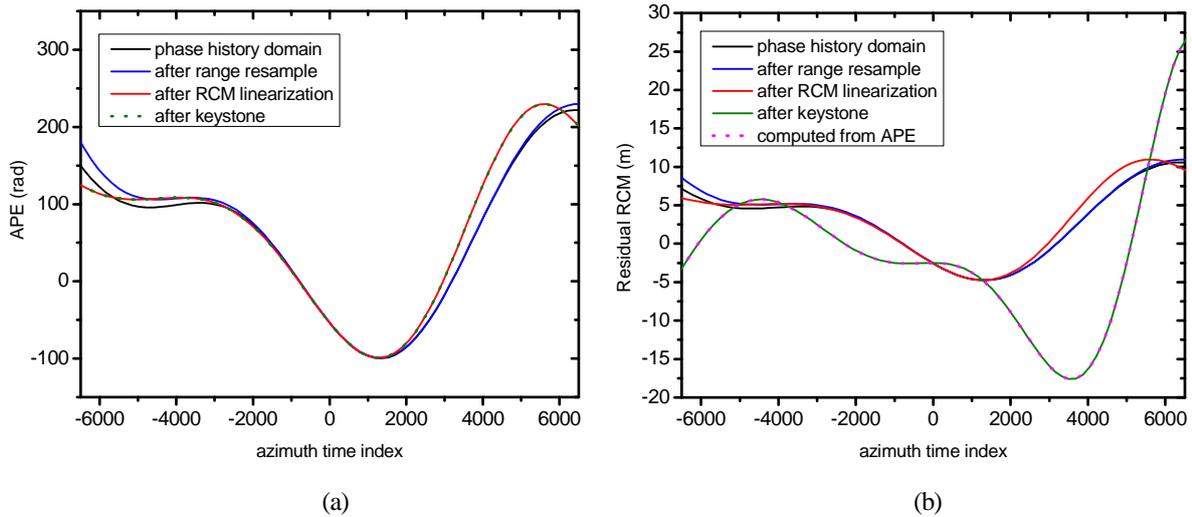

Fig.11. Illustration of (a) APE and (b) residual RCM in different processing stages of PFA.

*B. Real Data Results*

Real data collected by an experimental airborne SAR is applied to demonstrate the effectiveness of the proposed autofocus methods. The experimental radar operates in spotlight mode at X-band. Its transmitted signal has a bandwidth of 1.2GHz, corresponding to 0.13m theoretical range resolution. The processed synthetic aperture length is about 2300m, hence the nominal azimuth resolution would be 0.067m. Because no motion sensor data is available in hand, we assume that the radar platform flies at a constant velocity in image formation. Fig.12 (a) and (b) give the range compressed image (to show residual RCM clearly, magnified local image including strong scatterers is presented) and full compressed image, respectively, using PFA processing without any autofocus procedure applied. Although deterministic range migration has been compensated by polar format transformation, residual RCM is still large enough to exceed several range resolution cells. It can be clearly seen that the image suffers from severely 2-D defocus, i.e., the image exhibits smearing not only in the azimuth direction, but also in the range direction due to residual RCM. In this situation, conventional autofocus algorithm can't completely compensate for the errors. Residual degradation in image is still substantial. This can be clearly seen in Fig.12(c), which is produced by PFA with PGA applied.

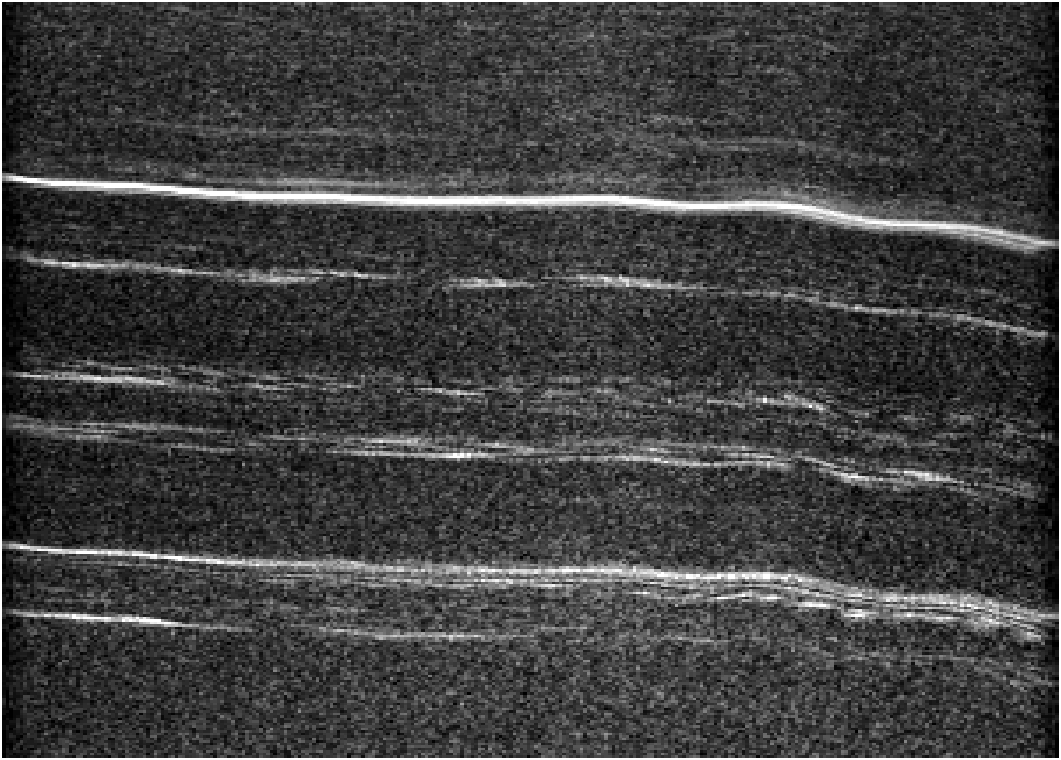

(a)

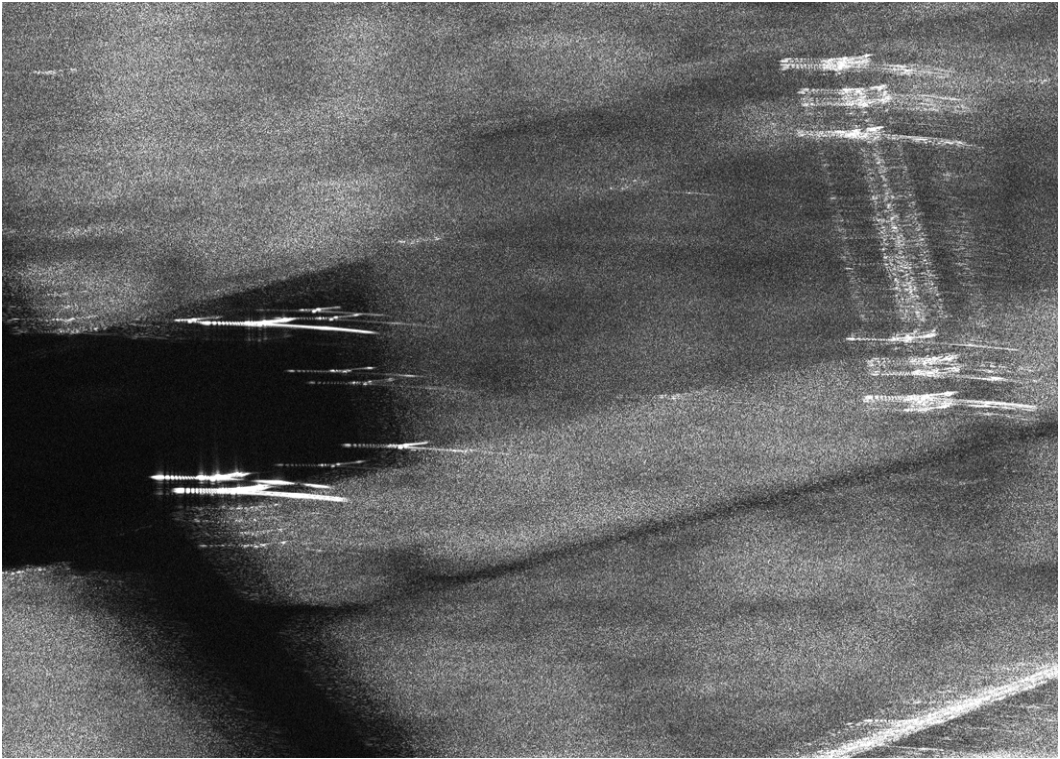

(b)

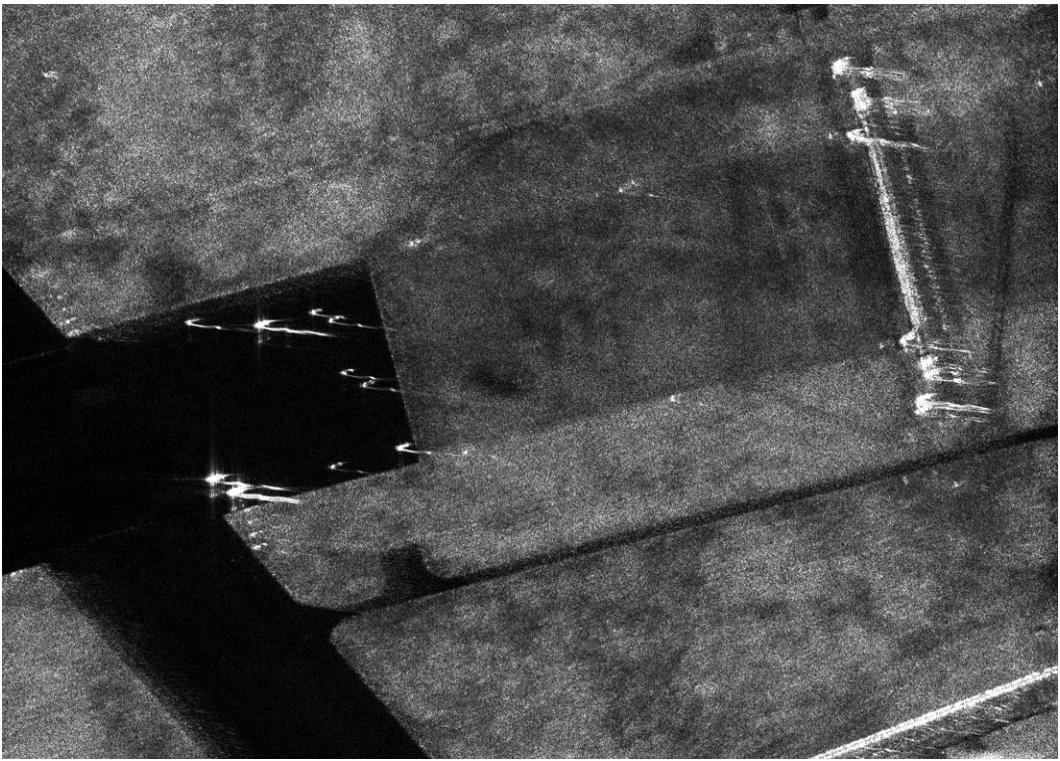

(c)

Fig.12. Images produced by PFA (a) range compressed image; (b) full compressed image; (c) refocused image by PGA.

To get a well focused image, both of the APE and residual RCM should be eliminated. For the APE estimation, we use the proposed subaperture autofocus approach in this example since the residual RCM is very large. The estimated APE is shown in Fig.13 (a). Based on this APE estimate, residual RCM is directly computed by using (28), and is presented in Fig.13 (b). The result also shows that low-pass filtering of the APE is necessary before using it to compute residual RCM.

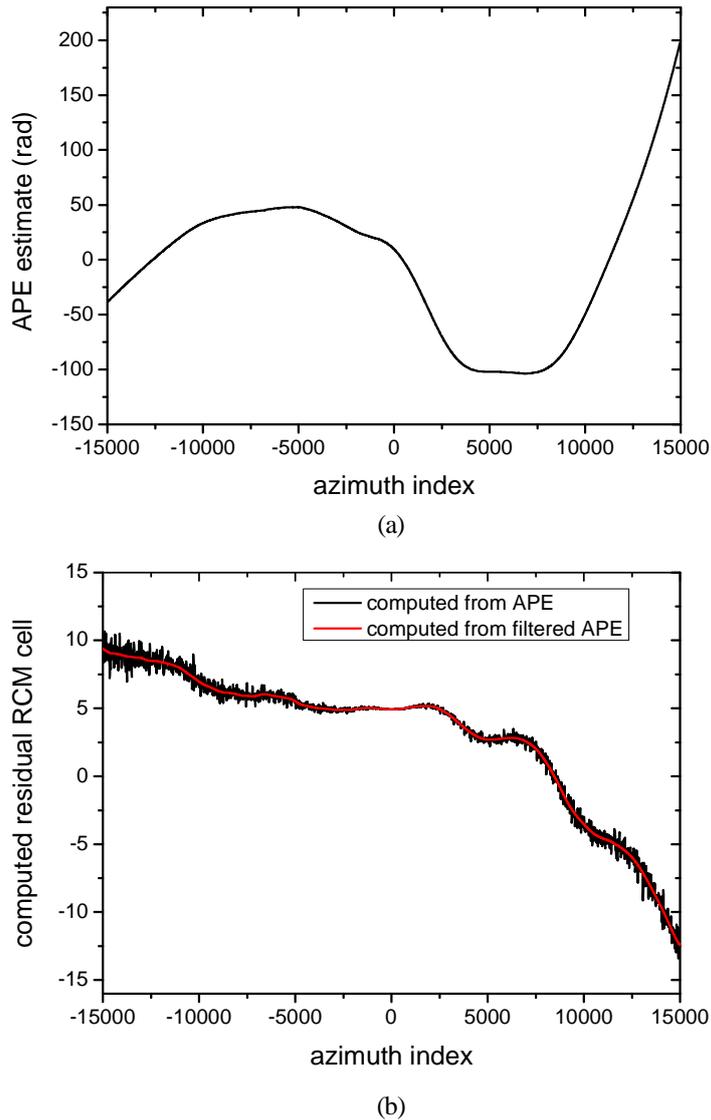

Fig.13. Error extracted from the PFA image. (a) APE estimate; (b) residual RCM computed from APE estimate.

Applying the above estimated APE and residual RCM to compensate for the defocused image, we get the range compressed image and the full compressed image, which are shown in Fig.14 (a) and (b), respectively. The reconstructions exhibit excellent quality.

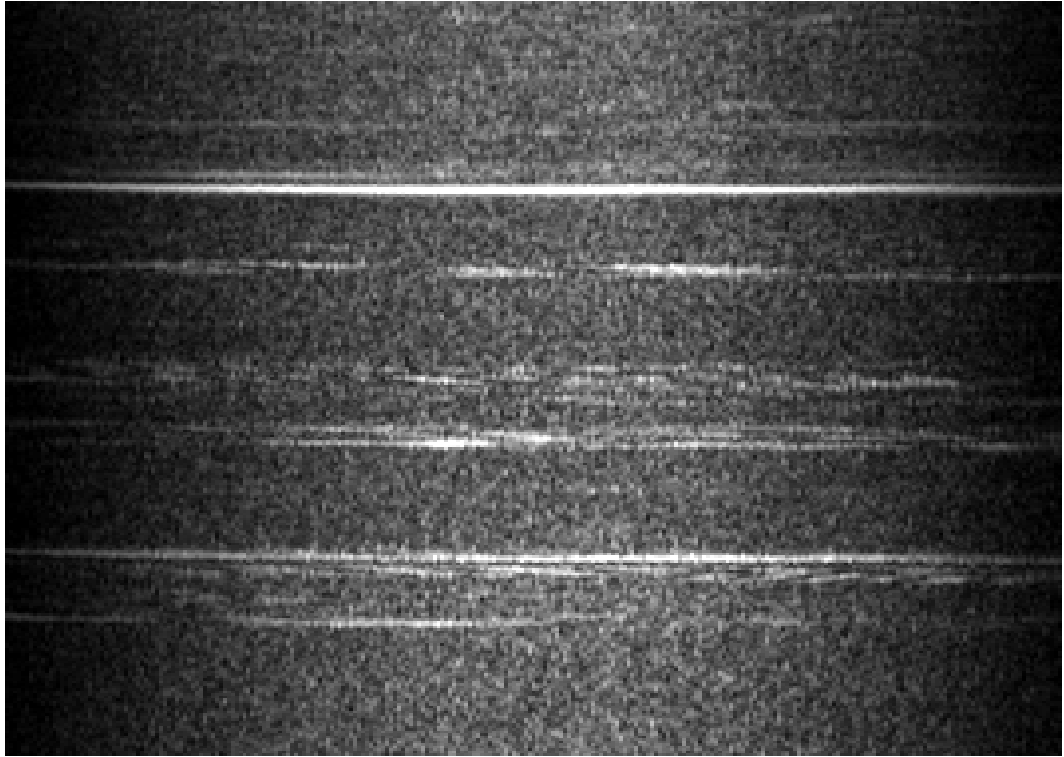

(a)

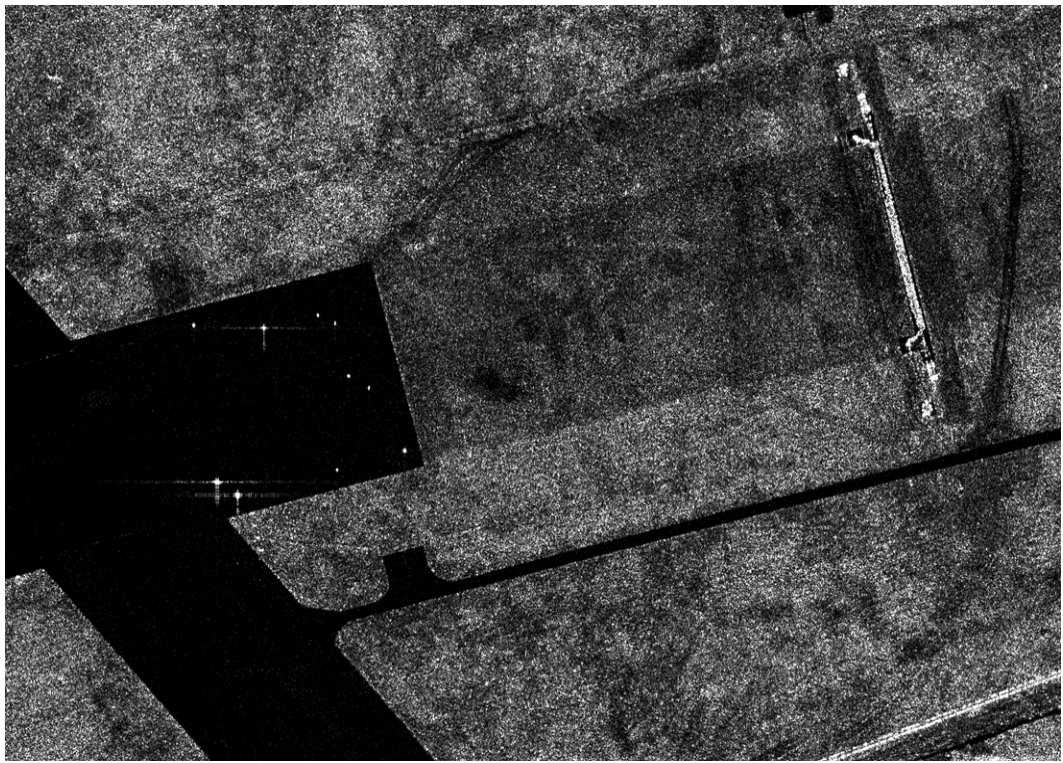

(b)

Fig.14. Corrected image by the proposed method. (a) range compressed image; (b) full compressed image.

## VI. Conclusions

Conventional autofocus algorithms apply one-dimensional azimuth phase error correction, while ignoring the effect of residual range cell migration, to the degraded image to refocus the targets. As the resolution increases, however, residual range cell migration exceeding a range resolution cell would become commonplace, or even inevitable. This makes the autofocus technique become inherently two-dimensional. In this work, the analytical relationship between APE and residual RCM in the framework of PFA is investigated. Based on this derived relationship, a 2-D autofocus scheme, which can compensate for the APE and residual RCM simultaneously, is proposed. The new approach possesses two obvious merits. First, its computational efficiency is relatively high since the residual RCM can be directly computed instead of additional estimation process. Second, the accurate estimator of APE obtained by existing autofocus method can make sure that the computed residual RCM has sufficient accuracy.

Nevertheless, the analysis and correction of residual errors in this work operate only in the framework of PFA. They can't be directly applied to other image formation algorithms, such as CSA and RMA, since different image formation algorithms have different effects on the uncompensated motion errors. However, it should be possible to derive the analytical relationship between APE and residual RCM after image formation by other algorithms, and this will be our future effort.